\documentstyle[12pt]{article}

\def\beqa{\begin{eqnarray}}
\def\eeqa{\end{eqnarray}}
\def\beq{\begin{equation}}
\def\eeq{\end{equation}}

\def\dmunu{_{\mu\nu}}

\def\uab{^{\alpha\beta}}

\def\pa{\partial}

\def\bib#1{$^{\ref{#1}}$}

\let\alp=\alpha

\def\pr{{\it Phys. Rev.}\ }
\def\prl{{\it Phys. Rev. Lett.}\ }

\def\cqg{{\it Class. Quantum Grav.}\ }

\def\apj{{\it Ap. J.}\ }
\def\aa{{\it Astron. Astrophys.}\ }

\def\prep{{\it Phys. Rep.}\ }

\begin{document}
\def\bib#1{[{\ref{#1}}]}
\begin{titlepage}
	 \title{ Gravitational Waveguides in Cosmology}

\author{{S. Capozziello$^{1,2}$, R. de Ritis$^{1,2}$, V. Mank'o$^{3,4}$,
A.A. Marino$^{1,2}$, G. Marmo$^{1,2}$}
\\ {\em $^{1}$Dipartimento di Scienze Fisiche, Universit\`{a} di Napoli,}
\\ {\em $^{2}$Istituto Nazionale di Fisica Nucleare, Sezione di Napoli,}\\
   {\em Mostra d'Oltremare pad. 19 I-80125 Napoli, Italy,}\\
 {\em $^{3}$Osservatorio Astronomico di Capodimonte,}\\
{\em Via Moiariello 16 I-80131 Napoli, Italy}\\
{\em $^{4}$Lebedev Physical Institute, Leninsky Pr., 53, Moscow 117924, 
Russia.}}

	      \date{}
	      \maketitle

	\begin{abstract}
We discuss the possibility that, besides the usual gravitational 
lensing, there may exist a sort of gravitational waveguiding in
cosmology which could explain some anomalous phenomena which 
cannot be understood by  the current  
 gravitational lensing models as the existence of
"brothers" objects having different brilliancy but similar spectra
and redshifts posed on the sky with large angular distance. 
Furthermore, such a phenomena could explain the huge luminosities
coming from quasars using the cosmological structures as selfoc--type
or planar waveguide.
We describe the gravitational waveguide theory and then we
discuss possible realizations in cosmology.  

 \end{abstract}       

\vspace{20. mm}
PACS: 98.80. Dr; 95.30. Sf

	      \vfill
	      \end{titlepage}

\section{\normalsize \bf Introduction}
Gravitational lensing has recently become a fundamental tool to investigate
large scale structure of the universe and to test cosmological models
\bib{peebles}.
One of the most interesting characteristics of gravitational lensing is 
that it acts on all scales. 
In fact, it provides a great amount of cosmological and astrophysical
applications like the determination of the Hubble parameter $H_{0}$ {\it via}
the measurement of  time delay $\Delta t$ between the observed lightcurves
of multiply imaged extragalactic sources \bib{borgeest},\bib{frieman};
the possibility of weighing the mass and describing the potential of lensing 
galaxies and galaxy clusters from the observation of multiply imaged
quasars, arcs and arclets \bib{fort},\bib{broadhurst}.
Furthermore, the gravitational lensing
plays a leading role in searching for dark matter since the frequency of
multiply imaged sources (e.g. quasars) depends on the cosmological density 
parameter $\Omega_{L}$ of compact objects. In fact a way to detect compact
objects in the universe with masses in the range 
$10^{-5}M_{\odot}-100M_{\odot}$ is based upon the detection of lensing 
effects which produce characteristic light variations of distant
compact sources. Particularly promising are the multiply
macro--imaged quasars whose lensing galaxy should have a large optical depth
for lensing effects \bib{paczynski},\bib{kaiser}
\bib{walsh} (at least 20 objects  are
identified with definite action of gravitational lensing; see, for 
example,
\bib{blioh},\bib{schneider}, \bib{ehlers}). 

The above kinds of analysis are possible if we have a model 
explaining the way of forming images such as the above--mentioned arcs, 
rings or simply double images and predicting the effects of the deflector 
\bib{bartelmann},\bib{kormann}.

From a theoretical point of view, lensing must be treated studying the 
geometry of the system source--lens--observer. This study is simple if we 
suppose that these are three points on a plane as well as 
if we consider  thin 
lens approximation: 
such hypotheses  are reasonable because of the large distances 
considered. A theoretical model  can be worked out by giving a 
specified form to the lens density, i.e. fixing its structure. From the 
density function, using the equations derived from the geometry, we can have 
predictions for the observed deviation of the source light and magnitude 
of every image.
A fundamental issue is how gravitational lensing
effects are connected with a theory of large scale structure formation;
in other words, how
 a theory of galaxy formation can directly furnish
gravitational lensing models.

It is well known that the gravitational
lensing may be explained using the action of a gravitational
field on the electromagnetic waves. In this case, the action of a media with
corresponding refraction index is,  for weak field approximation, 
completely determined by the Newtonian gravitational potential
which deflects and focuses the light rays.

In optics, however, there exist 
other types of devices, like optical fibers and  waveguides which use the same
deflection phenomena. The analogy with the action of a gravitational field
onto light rays may be extended to incorporate these other 
structures on the light. Shortly, it is 
 possible to suppose the existence of a sort of
gravitational waveguiding  \bib{dodma},\bib{dodonov}, \bib{dodma1}. 
But  this remark has not been studied, till now,  neither theoretically
nor experimentally. On the other hand, structures
like cosmic strings, texture and domain walls,
 which are produced at phase transiton in inflationary
models, can evolve into today observed filaments, clusters and groups
of galaxies and behave in a variety of ways with respect to
gravitational lensing effects.
In fact, the lensing by cosmic string was suggested as explaination of
the observation \bib{turner} of twins objects with anomally large angular
distance between the partners \bib{vilenkin},\bib{gott}.

The aim of this work is to discuss the properties of possible waveguides
in the universe and to suggest the explaination of some phenomena,
like quasar huge luminosities and large angular distances between
twins, as a by--product of their existence.
For example, a filament of galaxies can be considered  a sort of
waveguide preserving total luminosity of a source, if we have locally an
effective
gravitational potential of the form $\Phi(\bf{r})\sim r^{2}$, while
the planar structures generated by the motion of cosmic strings (the so 
called "wakes") can yield cosmological structures where the total
flux of light is preserved and the brightness of objects at high redshift,
whose radiation passes through such structures, appears higher 
to a far observer.

Sect. 2 is devoted to the discussion of the gravitational potential
intended as the refraction index of geometrical optics. 

In Sect. 3, we costruct the optical waveguide model using the 
paraxial approximation (the so called Fock--Leontovich approximation 
\bib{fock}).
We propose a model in which we consider, instead of a simple
gravitational lensing effect, the effect of a sort of system of lenses
which, combined in files or in planes, results as a waveguide.
In Sect. 4, we discuss the eventual cosmological realization of
such structures and the connection
with observations. Conclusions are drawn in Sect.5.

\section{\normalsize \bf  The propagation of light in a weak gravitational 
field}
It is possible to discuss the gravitational waveguide properties
\bib{dodma},\bib{dodonov},\bib{dodma1}  using
the electromagnetic field theory in a gravitational field described by
the metric tensor $g\dmunu$ \bib{blioh},\bib{ehlers}. In this context,
the behaviour of the electromagnetic field without sources in the presence of
a gravitational field may be described by the Maxwell equations 
\bib{skrotsky},\bib{plebansky}
\beq
\label{14}
\frac{\pa F_{\alp\beta}}{\pa x^{\gamma}}+
\frac{\pa F_{\beta\gamma}}{\pa x^{\alp}}+
\frac{\pa F_{\gamma\alpha}}{\pa x^{\beta}}=0\;;
\eeq
\beq
\label{15}
\frac{1}{\sqrt{-g}}
\frac{\pa}{\pa x^{\beta}}\left(\sqrt{-g} F^{\alpha\beta}\right)=0\,,
\eeq
where $F\uab$ is the electromagnetic field tensor and $g$ is the determinant
of the four--dimensional metric tensor. For a static gravitational field,
these equations can be reduced to the usual Maxwell equations describing the
electromagnetic field in media where the dielectric and magnetic tensor 
permeabilities are connected with the metric tensor $g\dmunu$ by the
equation \bib{landau}
\beq
\label{16}
\varepsilon_{ik}=\mu_{ik}=-g_{00}^{-1/2}[\mbox{det}g_{ik}]^{-1/2}g_{ik}\;;
\;\;\;\;\;i,k=1,2,3\,.
\eeq
If one has an isotropic model, the metric tensor is diagonal and 
the refraction index of "media" may be introduced by mimicking by the 
gravitational field
\beq
\label{17}
n({\bf r})=(\varepsilon\mu)^{1/2}\,,
\eeq
(it is worthwhile to note that such a situation can be
easily reproduced in cosmology \bib{ehlers}).

Following the standard procedure for deriving the scalar Helmholtz
equation \bib{ehlers} for the components of the electromagntetic field from 
the first order Maxwell equations, we get (for some arbitrary monochromatic
component of the electric field)
\beq
\label{18}
\frac{\pa^{2}E}{\pa z^{2}}+\frac{\pa^{2}E}{\pa x^{2}}
+\frac{\pa^{2}E}{\pa y^{2}}+k^{2}n^{2}({\bf r})E=0\,,
\eeq
where $k$ is the wave number.
 This procedure works if the relative change of
diffraction index on distances of the light wavelength 
 is small. 

The coordinate $z$, in Eq.(\ref{18}), is considered as the longitudinal one
and it can measure the space distance along the gravitational
field structure produced by a mass distribution
with an optical axis. Such a coordinate may also correspond to a distance
along the light path inside a planar gravitational field structure
produced by a planar matter--energy distribution in some regions
of the universe (like a string wake or a wall of galaxies 
see the discussion in Sect. 4).
For weak gravitational fields, considered also to describe usual gravitational 
lensing effects, the metric tensor components are expressed in terms of
the Newton gravitational potential $\Phi$ as 
\bib{blioh},\bib{ehlers},\bib{landau}
\beq
\label{19}
g_{00}\simeq 1+2\frac{ \Phi ({\bf r})}{c^{2}}\;;\;\;\;\;\;
g_{ik}\simeq -\delta_{ik}\left(1-2\frac{\Phi({\bf r})}{c^{2}}\right)\;;
\eeq
where we are assuming the approximation
$\Phi/c^{2}\ll 1$. Then, due to relations (\ref{16}) and (\ref{19}), the 
refraction index $n({\bf r})$ in (\ref{17}) can be expressed in terms of the 
gravitational potential $\Phi({\bf r})$ produced by some matter distribution.
Such a weak field situation is realized for  cosmological structures
which give rise to the gravitational lensing effects connected to several
observable phenomena (multiple images, magnification, image distorsion,
arcs and arclets); with some cautions, we can use the same scheme also in
strong lensing approximation
\bib{ehlers}. Here, we are interested to a specific
application which, we believe, is realized by some kinds of gravitational
systems: is it possible to realize cosmological waveguide effects considering
string--like or planar--like distribution of matter? Are such effects 
observable? In the next section we construct a gravitational waveguide
model using the above hypotheses on gravitational 
field and light propagation.
After we discuss some practical cosmological applications.

\section{\normalsize \bf  The gravitational waveguide model}
If one has a matter distribution with some axis like a cylinder with  dust
or like a planar slab with dust, it is possible to consider the 
electromagnetic field radiation propagating paraxially. The parabolic
approximation \bib{fock} is used for describing light propagation
in media and in devices as optical fibers \bib{lifsits}.
Below, we will discuss the possibility to use this approximation for
describing electromagnetic radiation propagating in a weak gravitational
field.

Let us consider, the scalar equation (\ref{18}) and the electric field $E$
of the form 
\beq
\label{1}
E=n_{0}^{-1/2}\Psi\exp\left(ik\int^{z}n_{0}(z')dz'\right)\;;
 \;\;\;\;\;n_{0}\equiv n(0,0,z)\,,
\eeq
where $\Psi(x,y,z)$ is a slowly varying spatial amplitude along the
$z$ axis,
and $exp(iknz)$ is a rapidly oscillating phase factor.
Its clear that the beam propagation  is along the $z$ axis. We rewrite 
Eq.(\ref{18})  neglecting second order derivative in longitudinal
coordinate $z$ and obtain a Schr\"odinger--like equation for $\Psi$:
\beq
\label{2}
i\lambda\frac{\pa \Psi}{\pa \xi}=-\frac{\lambda^{2}}{2}
\left(\frac{\pa^{2}\Psi}{\pa x^{2}}+\frac{\pa^{2}\Psi}{\pa y^{2}}\right)
+\frac{1}{2}\left[n_{0}^{2}(z)-n^{2}(x,y,z)\right]\Psi\,,
\eeq
where $\lambda$ is the electromagnetic radiation wavelength and we adopt
the new variable
\beq
\label{3}
\xi=\int^{z}\frac{dz'}{n_{0}(z')}\,,
\eeq
normalized with respect to the refraction index \bib{dodma}
(for our application, $n_{0}(z)\simeq 1$ so that $\xi$ coincides essentially
with $z$).

At this point, it is worthwhile to note that
if one has the distribution of the matter in the form of cylinder with
a constant (dust) density $~\rho_{0} ~$,  the gravitational potential inside  
 has a  parabolic  profile providing waveguide effect 
for electromagnetic radiation analogous to sel--foc optical waveguides
realized in fiber optics. In this case, Schr\"odinger--like equation 
is that of two--dimensional quantum harmonic oscillator
for which the mode solutions exist in the form of Gauss--Hermite
polynomials (see, for example, \bib{manko}). In the case of inhomogeneous
longitudinal dust distribution in the cylinder (that is $~\rho \,(z)\,$),
 the 
Schr\"odinger-like equation describes the model of two-dimensional parametric
oscillator for which the mode solutions, in the form of modified Gaussian
and Gauss--Hermite polynomials, exist with parameters 
determined by the density dependence on longitudinal coordinate.

As a side remark, it is interesting to stress that,
considering again Eq.(\ref{2}), the term in square brackets in the rhs 
plays the role of the potential
in a usual Schr\"odinger equation; the role of Planck constant is
now assumed by $\lambda$. Since the refraction index can be expressed in terms
of the Newtonian potential when we consider the propagation of light in a 
gravitational field, we can write the  potential
in (\ref{2}) as
\beq
\label{pot}
U({\bf r})=\frac{2}{c^{2}}[\Phi(x,y,z)-\Phi(0,0,z)]\,.
\eeq
The waveguide effect depends specifically on the 
shape of potential (\ref{pot}): 
for example,  the radiation from a remote source
does not attenuate if $U\sim r^{2}$; this situation is realized 
supposing a "filamentary" or a "planar" mass distribution with constant
density $\rho$. Due to the Poisson equation, the potential inside the filament
is a quadratic function of the transverse coordinates, that is of
$r=\sqrt{x^{2}+y^{2}}$ in the case of the filament and of $r=x$ in the
case of the planar structure (obviously the light propagates in the
"remaining" coordinates: $z$ for the filament, $z,y$ for the plane).
In other words, if the radiation, travelling from some source, undergoes
a waveguide effect, it does not attenuate like $1/R^{2}$ as usual,
but it is, in some sense conserved; this fact means that the source
brightness will turn out to be much stronger than the brightness of analogous
objects located at the same distance (i.e. at the same redshift 
$Z$) and the apparent energy released by the source will be anomalously large.

To fix the ideas, let us estimate how the  electric field (\ref{1}) 
propagates
into an ideal  filament whose internal potential is
\beq
\label{internal}
U(r)=\frac{1}{2}\omega^{2}r^{2}\,,\;\;\;\;\;\;
\omega^{2}=\frac{4\pi G \rho}{c^{2}}
\eeq
where $\rho$ is constant and $G$ is the Newton constant.
 A spherical wave from a source, 
\beq
\label{sphere}
E=(1/R)\exp(ikR)\,, 
\eeq
can be represented
in the paraxial approximation as 
\beq
\label{4}
E(z,r)=\frac{1}{z}\exp\left(ikz+\frac{ikr^{2}}{2z}-
\frac{r^{2}}{2z^{2}}\right)\,,
\eeq
where we are using the expansion
\beq
\label{5}
R=\left(z^{2}+r^{2}\right)^{1/2}\approx 
z\left(1+\frac{r^{2}}{2z^{2}}\right)\,,\;\;\;\;r\ll z\,.
\eeq
It is realistic to  assume $n_{0}\simeq 1$ so that, from (\ref{3}), $\xi=z$.
Assume now that the starting point of the filament of length $L$ is at
a distance $l$ from a source shifted by a distance $a$ from the filament
axis in the $x$ direction. The amplitude  $\Psi$ of the field $E$, entering
the wave guide is
\beq
\label{6}
\Psi_{in}=\frac{1}{l}\exp
\left[\frac{ikl-1}{2l^{2}}\left((x-a)^{2}+y^{2}\right)\right]\,,
\eeq
and so in (\ref{sphere}), we have
$ R=\left(l^{2}+y^{2}+(x-a)^{2}\right)^{1/2}.$

We can calculate the amplitude of the field at the exit of the filament
by the equation
\beq
\label{8}
\Psi_{f}(x,y,l+L)=\int 
dx_{1}dy_{1}G(x,y,l+L,x_{1},y_{1},l)\Psi_{in}(x_{1},y_{1},l)\,,
\eeq
where $G$ is the Green function of Eq.(\ref{2}). For the potential 
(\ref{internal}), $G$ has the form
\beqa
\label{9}
G(x,y,l+L,x_{1},y_{1},l)&=&\frac{\omega}{2\pi i\lambda\sin\omega\lambda}
\nonumber\\
 &\times& \exp\left(\frac{i\omega[\cos\omega L
(x^{2}+y^{2}+x_{1}^{2}+y_{1}^{2})-2(xx_{1}+yy_{1})]}{2
\pi i\lambda\sin\omega\lambda}\right)\,,
\eeqa
which is the propagator of the harmonic oscillator. The integral
(\ref{8}) is Gaussian and can be exactly evaluated
\beqa
\label{10}
\Psi_{f}&=&\frac{\omega l}{\omega l^{2}\cos\omega L+(l+i\lambda)\sin\omega l}
\nonumber\\
&\times&\exp\left(-\frac{1}{2}\frac{(x^{2}+y^{2})
[(\omega l k)^{2}-\omega k(i+k l)\cot\omega l]-a^{2}\omega k(i+k l)\cot\omega
L}{1-ikl-ik\omega l^{2}\cot\omega l}\right)\nonumber\\
&\times&\exp
\left(-\frac{2xa\omega k(1+kl)}{2\sin\omega L(1-ikl-ik\omega l^{2}\cot\omega 
L)}\right)\,.
\eeqa
The parameter $l$ drops out of the denominator of the pre--exponential factor
if the length $L$ satisfies the condition
\beq
\label{11}
\tan\omega L=-\omega l\;.
\eeq
Eq.(\ref{10}) is interesting in two limits.
If $\omega l\ll 1$, we have
\beq
\label{112}
\Psi_{f}=\frac{1}{i\lambda}\exp
\left\{-\frac{l+i\lambda}{2\lambda^{2}l}\left[(x+a)^{2}+y^{2}\right]\right\}\,,
\eeq
which means that the radiation emerging from a point with coordinate
$(a,0,0)$ is focused near  a point with coordinates $(-a,0,l+L)$
(that is the radius has to be of the order of the wavelength). 
This means that, when the beam from an extended source is focused
inside the waveguide in such a way that, at a distance $L$, 
Eq.(\ref{11}) is satisfied,  an inverted image of the source is formed,
 having the very
same geometrical dimensions of the source. The waveguide "draws" the
source closer to the observer since, if the true distance of the observer
from the source is $R$, its image brightness will correspond to that of
a similar source at the closer distance
\beq
\label{eff}
R_{eff}=R-l-L\,.
\eeq
If we do not have $\omega l\ll 1$, we get (neglecting the term $i\lambda/l$
compared with unity)
\beq
\label{13}
\Psi_{f}=\frac{\sqrt{1+(\omega l)^{2}}}{i\lambda}\exp
\left\{-\frac{1+(\omega l)^{2}}{2\lambda^{2}}
\left[y^{2}+\left(x+\frac{a}{\sqrt{1+(\omega l)^{2}}}\right)^{2}
\right]\right\}\,,
\eeq
from which, in general, the size of the image is decreased by
a factor $\sqrt{1+(\omega l)^{1/2}}$. The amplitude increases by the
same factor, so that the brightness is $(R/R_{eff})$ times larger.

In the opposite limit $\omega l\gg 1$, we have 
$\tan\omega L\rightarrow\infty$, so that $L\simeq \pi/\omega$, that is
the shortest focal length of the waveguide is
\beq
\label{foc}
L_{foc}=\sqrt{\frac{\pi c^{2}}{4G\rho}}\,,
\eeq
which is the length of focusing of the initial beam of light trapped by
the gravitational waveguide.
All this arguments apply if the waveguide has (at least roughly) a
cylindrical geometry.
The theory of planar waveguide is similar but we have to consider only
$x$ as transverse dimension and not also $y$. 

The cosmological feasibility of a waveguide depends on the geometrical 
dimensions of the structures, on the connected densities
and on the limits of applicability of the above idealized scheme.
In the next section, we shall discuss these features and the possible
candidates which could give rise to observable effects.

\section{\normalsize \bf  Observable effects and 
cosmic structures as waveguides}
The gravitational waveguide effect has the same physical reason
that has the gravitational lenses effect which is the electromagnetic
wave deflection by the gravitational field (equivalent to the deflection
of light by a refractive media). However, 
there are essential differences producing
specific predictions for observing the waveguide effect. The gravitational 
lenses are usually considered as compact objects with strong enough
gravitational potential. The light rays deflected by gravitational lenses
move outside of the matter which forms the gravitational lens itself. 
The gravitational waveguide as well as optical waveguide is noncompact
long structure which may contain small matter density and deflection
of the light by each element of the structure is very small. Due
to very large scale sizes of the structure (we give an extimation below),
 the electromagnetic 
radiation deflection by the gravitational waveguide occurs and,
in principle, it may be observed. We will mention, for example, a possibility 
of brilliancy magnification of the long distanced objects (like quasars)
with large red shift as consequence of the waveguiding structure 
existence between the object and the observer. This effect exists also 
for a gravitational lens located between the object and the observer,
but the long gravitational waveguide may give huge 
magnification, since the radiation propagates along the waveguide with
functional dependence of the intensity on the distance which does not 
decrease as
$~\sim 1/R^2~$, characteristic for free propagation. The gravitational lens,
being a compact object, collects much less light by deflecting the rays
to observer than the gravitational waveguide structure transporting to 
the direction of observer all trapped energy (of course, one needs to
take into account losses for scattering and absorbtion). From that
point of view, it is possible that enormous amount of radiation emitted by
quasars is only seemingly existing. The object may radiate a resonable 
amount of energy but the existing waveguide structure transmit the energy
in high portion to the observer. Similar ideas, related to gravitational 
lensing, were discussed in \bib{barn} but, since above mentioned reasons,
the singular lens or even few  aligned strong lenses cannot produce
effect of many orders of magnitude magnification of brilliancy. 
The waveguide effect may explain the anomalous high luminosity
observed in  quasars. In fact, quasars are objects at very high
redshift which appear almost as point sources but have luminosity
that are about one hundred times than that of a giant elliptical galaxy
(quasars have luminosity which range between $10^{38}-10^{41}$ W).
For example,
PKS 2000-330 has one of the largest known redshifts $(Z=3.78)$
with a luminosity of $10^{40}$ W.
Such a redshift corresponds to a distance of $3700$ Mpc, 
if it is assumed that its origin is due to
the expansion of the universe and the Hubble constant is assumed
$H=75$km s$^{-1}$ Mpc. This means that light left the quasar when
the size of the universe was one--fifth of its present age where
no ordinary galaxies (included the super giant radio--galaxies)
are observed. 
The quasars, often, have both emission and absorption lines in their spectra.
The emission spectrum is thought to be produced in the quasar itself;
the absorption spectrum, in gas clouds that have either been ejected from
the quasar or just happen to lie along the same line of sight.
The brightness of quasars may vary rapidly, within a few days or less. Thus,
the emitting region can be no larger than a few light--days, i.e.
about one hundred astronomic units. This fact excludes that quasars could
be galaxies (also if most astronomers think that quasars are extremely
active galactic nuclei).

The main question is how to connect this small size with the so high
redshift and luminosity. 
For example, H.C. Arp discovered small systems of quasars and galaxies where
some of the components have widely discrepant redshifts \bib{arp}.
For this reason, quasar high redshift could be produced by some unknown 
process and  not being simply of cosmological origin. This claim is very
controversial. However there is a fairly widely accepted
preliminary model which, in principle,  could unify all the forms
of activities in galaxies (Seyfert, radio, Markarian galaxies and
BL Lac objects). According to this model, most galaxies contain a compact
central  nucleus with mass $10^{7}-10^{9}$ M$_{\odot}$ and diameter $< 1 $ 
pc. For some reason,
the nucleus may, some times, release an amount of energy exceeding 
the power of all the rest of the galaxy. If there is only little gas
near the nucleus, this leads to a sort of double radio source. If the nucleus 
contains much gas, the energy is directly released as radiation and one
obtains a Seyfert galaxy or, if the luminosity is even larger, a quasar.
In fact, the brightest type 1 Seyfert galaxies and faintest quasars are
not essentially different in luminosity ($\sim 10^{38}$ W) also if the
question of redshift has to be  explained (in fact quasar are, apparently,
much more distant). Finally, if there is no gas at all near such an active 
nucleus, one gets BL Lac objects. These objects are similar to quasar
but show no emission lines. However the mechanism to release  such a large
amount of energy from active nuclei or quasars is still unknown. 
Some people suppose that it is connected to the releasing
of gravitational energy due to the interactions of internal components
of quasars. This mechanism is extremely more efficient than the
releasing of energy during the ordinary nuclear reactions. The necessary
gravitational  energy could be produced, for example, 
as consequence of the falling
of gas in a very deep potential well as that connected with a very
massive black hole. Only with this assumption, it is possible 
to justify a huge luminosity, a cosmological
 redshift and a small size for the quasars.

An alternative explaination could come from waveguiding effects. 
As we have discussed, if light travels within a filamentary or a
planar structure, whose Newtonian gravitational potential is quadratic
in the transverse coordinates, the radiation is not attenuated, moreover
the source brightness is stronger than the brightness of analogous
object located at the same distance (that is at the same redshift). In other
words, if the light of a quasar undergoes a waveguiding effect, 
due to some structure
along the path between it and us, the apparent energy released by the
source will be anomalously large, as the object were at a distance
(\ref{eff}). Furthermore, if the approximation $\omega l\ll 1$
does not hold,  the dimensions of resulting image would 
be decreased by a factor
$\sqrt{1+(\omega l)^{2}}$ while the brightness would be $(R/R_{eff})^{2}$,
larger, then explaining how it is possible to obtain so large emission by
such (apparently)  small  objects. In conclusion, the existence of a 
waveguiding 
effect may
prevents to take into consideration exotic mechanism 
in order to produce huge amounts of energy
(as the existence of a massive black hole inside a galactic core) and it
may justify why it is possible to observe so distant objects of small
geometrical size.

Another effect concerning the quasars may be directly connected with
multiple images in lensing. The waveguide effect does not disappear
if the axis of "filament" or if the guide plane is bent smoothly in space.
As in the case of gravitational lenses, we can observe "twin" images if
part of the radiation comes to the observer directly from the source, and 
another part is captured by the bent waveguide. The "virtual" image
can then turn out to be brighter than the "real" one (in this case we may
deal with "brothers" rather than "twins" since parameters like,
spectra, emission periods and chemical compositions are similar but
the brightnesses are different). Furthermore, such a bending in waveguide
could explain large angular separations among the images of the same
object which cannot be explained by the current lens models (pointlike
lens, thin lens and so on). 
  
Now the issue is: are there  cosmic structures which can furnish
workable models for waveguides? Have they to be "permanent" structures
or may the waveguide effect be accidental (for example an alignment of 
galaxies
of similar density and structure, due to cosmic shear and inhomogeneity, 
may be available as waveguide just for a limited interval of time
\bib{valdes})? In general both points of view may be reasonable and 
here we will outline both of them.
Furthermore we have to consider the problem 
of the abundance of such structures: are they  common and everywhere in
the universe or are they peculiar and 
located in particular regions (and eras)?

We have to do a first remark on the densities of waveguide structures
which allow observable effects \bib{dodma}.
Considering Eq.(\ref{foc}) and introducing into it the critical
density of the universe $\rho_{c}\sim 10^{-29}$ g/cm$^{3}$ (that is the
value for which the density parameter is $\Omega=1$), we obtain
$L_{foc}\sim 5\times 10^{4}$ Mpc which is an order of magnitude larger than
the observable universe and it is completely unrealistic.  
On the contrary, considering a  typical galactic
density as $\rho\sim 10^{-24}$ g/cm$^{3}$, we obtain $L_{foc}\sim 100$ Mpc,
which is a typical size of large scale structure (e.g.
the Great Wall  has such dimensions  and also a filament
of galaxies can have such a length \bib{huchra}). 

However, an important issue has to be 
taken into consideration: the absorption and the scattering of light by 
 the matter inside the filament or the planar structure increase with 
density and at certain crital value the waveguide effect can be invalidated
\bib{dodma}. 
For the smaller frequency 
of broadcast range (due to strong dependence of the absorbtion cross
section on the electromagnetic wavelength) $~\sigma \sim \sigma _T\,
(\omega /\omega _0)^4,$ where Thomson cross-section $~\sigma _T=6\cdot
10^{-25}~\mbox {cm}^2~$ and the characteristic atomic frequency is
$~\omega _0\sim 10^{16}~\mbox {s}^{-1},$ the ratio 
$~\omega /\omega _0\ll 1\,,$ and the absorbtion is small. It means that
the absorbtion length $~L_a=m_p/\rho \,\sigma \,,$ (where the mass of
proton $~m_p~$ is approximately equal to the hydrogen atom mass) is
larger than the focusing length $~L_a<L_{foc}$ for the electomagnetic waves 
of broadcast range. Thus, the magnification of electromagnetic waves
may be not masked by essential energy losses due to light absorbtion and 
scattering processes.
However, no restrictions exist practically  if the radio band
 and a thickness of the structure $r>10^{14}$cm are considered.

In such a case, the 
relative density change 
between the background and the structure density
is valid till $\delta \rho/\rho\leq 1$ . This means that we have
to stay in a linear perturbation density regime.

By such hypotheses, practically all the observed large scale structures
like filaments, walls, bubbles and clusters of galaxies can result as
candidates for waveguiding effect if the restrictions on density,
potential and waveband are respected (in optical band, such  phenomena
are possible but the density has to be chosen with some care). 

Also primordial structures (produced in inflationary phase transition
and surviving later),
like cosmic string, could furnish waveguides. In fact, in 
weak energy limit approximation, such objects are internally described
by the Poisson equation $\nabla^{2} \Phi=\rho_{0}$ and externally by 
$\nabla^{2}\Phi=0$ and, furthermore,
they act as gravitational lenses after the formation of the quasar
\bib{vilenkin},\bib{gott}. It is easy to recover an internal 
potential of the form $\Phi\sim r^{2}$ and, considering the dynamical
evolution after the decoupling, lengths in the required ranges for waveguide
(e.g. $\sim 100$ Mpc). The main problem is due to the fact that also
after the evolution to macroscopic sizes, strings remain "wires" without
becoming cylinders, that is their thickness remain well below 
$r\sim 10^{14}$ cm, the minimal value required to get observable effects.
However, we have not considered the scaling solutions (see for example
\bib{kibble}) from which such wires could evolve in cylindrical structures
(with transverse sizes non trivial with respect to the background).

Other two interesting features are connected with cosmic strings:
the first is that their motion with respect to the background
produces wakes and filaments which, later, are able to  evolve in large
scale structures systems of galaxies \bib{vachaspati}. For example, at 
decoupling $(Z\sim 1000)$,
a string can produce a wake, which consists in a planar structure,
with side $\Delta r\sim 1$ Mpc and constant surface density
$\sigma_{0}\sim 3\times 10^{11}$M$_{\odot}$Mpc$^{-2}$.
Such a feature is interesting for large scale structure formation
and can yield a planar waveguide with today observable effects.
The second fact is that inflationary phase transition can produce
a large amount of cosmic strings which, evolving, can give rise to
a string network pervading all the universe \bib{peebles},\bib{vilenkin}.
In such a case, if they evolve in cylindrical or planar structure, we may
expect large probabilities to detect waveguiding effects.

Concerning the second point of view (that is the existence of temporary
waveguiding effects), it could be related to the observation of
objects possessing anomalously large (compared with their neighbours)
angular motion velocities (an analysis in this sense
could come out in mapping galaxies with respect to their redshift and proper 
velocities, see for example \bib{davis}). 
Such a phenomenon could mean that one observes
not the object itself, but its image transmitted through the moving 
gravitational waveguide. The waveguide itself could change its form or
it could be due to temporary alignments of lens galaxies.
In this case, the image of the object could move with essentially different
angular velocity than that of the observable neighbour objects whose light
reaches the observer directly (not throught the waveguide).
The discovery of long distanced objects with anomalous velocity
(and brightness) could support the hypothesis of gravitational waveguide
effect, while the evolution of the waveguide, its destruction or change
of the axis direction (from the orientation to the Earth) could
produce the effect of the disappearence (or the appearence) of
the observed object. For this analysis, it is crucial to consider
long period astronomical observations and deep pencil beam surveys of
galaxies and quasars.

\section{\normalsize \bf  Conclusions}
In this paper, we have discussed the possible existence of gravitational
waveguide effects in the universe and constructed a radiation propagation
model to realize them. As in the case of gravitational lensing, several
phenomena and cosmic structures could confirm their existence, starting from
primordial object like cosmic strings to temporary alignement of
evolved late--type galaxies. Furthermore, due to the wavelength considered,
they could give observable effects in optic, radio or microwave bands or,
alternatively, considering the propagation of other weak interacting
particles as the neutrinos. The experimental feasibility for
the detection could have serious troubles due to the need of long
period observations or due to the discrimination among data coming from
objects which have undergone waveguide effects and objects which not.

In any case, if such a hypothesis will be confirmed in some of the
above quoted senses, we shall need a profound revision of our
conceptions of large scale structure  and matter distribution.

Finally we want to stress that our treatment does not concern only 
electromagnetic radiation:
actually 
a waveguide effect could be observed
 also for streams of neutrinos or other
particles which gravitationally interact with the filament (or the plane),
in this sense it could result useful also in other fields of
astrophysics and fundamental physics.

\begin{center}
{\bf REFERENCES}
\end{center}

\begin{enumerate}
\item\label{peebles} 
P.J.E. Peebles  {\it Principles of 
Physical Cosmology} (Princeton Univ. Press.,
Princeton 1993)
\item\label{borgeest} 
U. Borgeest \aa {\bf 128} 162 (1983)
\item\label{frieman} 
J.A. Frieman, D.D. Harari, and G.C. Surpi
\pr {\bf D 50} 4895 (1994)
\item\label{fort} 
B. Fort, and Y. Mellier \aa {\it Rev.}  {\bf 5} 239 (1994)
\item\label{broadhurst} 
T.J. Broadhurst, A.N. Taylor, and J.A. Peakock
\apj {\bf 438} 49 (1995)
\item\label{paczynski} 
B. Paczynski {\it Gravitational Lenses} 
Lecture Notes in Physics {\bf 406}, p. 163, Springer--Verlag, Berlin (1992) 
\item\label{kaiser} 
R. Kaiser {\it Gravitational Lenses} 
Lecture Notes in Physics {\bf 404}, p. 143, Springer--Verlag, Berlin (1992) 
\item\label{bartelmann} 
M. Bartelmann and  P. Schneider \aa {\bf 268} 1 (1993)
\item\label{kormann} 
R. Kormann, P. Schneider, and M. Bartelmann
\aa {\bf 284}, 285 (1994)
\item\label{walsh}
D. Walsh, R.F. Carswell,  and R.J. Weymann {\it Nat.} {\bf 279}, 381 (1979)
\item\label{blioh}
P.V. Blioh and A.A. Minakov {\it Gravitational Lenses}
Kiev, Naukova Dumka (1989) (in Russian)
\item\label{schneider}
P. Schneider {\it Cosmological Applications of Gravitational Lensing},
Lecture Notes in Physics, eds. E. Martinez--Gonzales, J.L. Sanz,  
Springer Verlag, Berlin (1996), Astro--Ph/9512047 (1995).
\item\label{ehlers}
P.Schneider, J. Ehlers, and E.E. Falco {\it Gravitational Lenses}
Springer--Verlag, Berlin (1992).
\item\label{lifsits}
E. M. Lif\v{s}its, L.P. Pitaevskij {\it Elettrodinamica dei mezzi continui},
Ed. Riuniti, Roma (1986).
\item\label{dodma}
V.V. Dodonov and  V.I. Man'ko, {\it Gravitational waveguide},
Preprint of the  Lebedev Physical Institute Proceedings, No. 255
(Moscow, 1988); {\it J. Soviet Laser Research} (Plenum Press), {\bf 10},
240 (1989); {\it Invariants and Evolution of Nonstationary Quantum Systems} 
Proceedings of the Lebedev Physical Institute, {\bf 183}
 Nova Science Publishers N.Y. (1989).
\item\label{dodonov}
V.V. Dodonov in:
{\it Proceedings of the First International Sakharov Conference} 
Moskow May 21--25 (1992), p. 241
Edt. L. V. Keldysh, V. Ya. Fainberg, Nova Science Publishers N.Y. (1993)
\item\label{dodma1}
V.V. Dodonov, O.V. Man'ko, and V.I. Man'ko, in
{\it Sqeezed and Correlated States of Quantum Systems} 
Proceedings of the Lebedev Physical Institute, {\bf 205} p. 217
 Nova Science Publishers N.Y. (1993).
\item\label{turner} 
E. L. Turner, D. P. Schneider, B. F. Burke, J. N. Hewitt,
G. L. Langston, J. E. Gunn, C. R. Lawrence, and M. Schmidt, {\sl Nature},
{\bf 321} 142 (1986).
\item\label{vilenkin}
A. Vilenkin, \apj {\bf 282}, L51 (1984); \prep {\bf 121}, 263 (1985);\\
\item\label{gott}
J.R. Gott III, \apj {\bf 288}, 422 (1985).
\item\label{fock} 
A. M. Leontovich and V. A. Fock, {\sl Zh. Eksp. Teor.
Fiz.}, {\bf 16}, 557 (1946).
\item\label{manko} 
V. I. Man'ko, in ~{\it Lee Methods in Optics}, Lecture 
Notes in Physics, Eds. S. Mondragon and K.-B. Wolf, {\bf 250} 193 (1986).         
\item\label{kolb} 
E.W. Kolb, M.S. Turner {\it The Early Universe}
(Addison--Wesley Pub. Co., Menlo Park 1990)
\item\label{binney}  
J. Binney,  S. Tremaine  {\it Galactic Dynamics} (Princeton
Univ. Press.: Princeton 1987)
\item\label{mihalas} 
D. Mihalas, J.J. Binney {\it Galactic Astronomy}
2nd. ed. San Francisco, Freeman (1981)
\item\label{refsdal} 
S. Refsdal, J. Surdej {\it Rep. Prog. Phys.} {\bf 57}, 117 (1994).
\item\label{landau}
L. D. Landau and E. M. Lif\v{s}its {\it The Classical Theory of Fields},
Pergamon Press, N.Y. (1975).
\item\label{skrotsky}
G.V. Skrotsky, {\it Doklady} AN SSSR, {\bf 114}, 73 (1957).
\item\label{plebansky}
J. Plebansky, \pr, {\bf 118}, 1396 (1960).
\item\label{arp}
H.C. Arp {\it Quasars, Redshifts and Controversies} Berkeley: Interstellar
Media, $\$ 7$ (1987).
\item\label{valdes}
F. Valdes, J.A. Tyson, J.F. Jarvis \apj {\bf 271}, 431 (1983).
\item\label{huchra}
V. de Lapparent, M.J. Geller, J.P. Huchra \apj {\bf 302}, L1 (1986).\\
M.J. Geller, J.P. Huchra {\it Science} {\bf 246}, 897 (1989). 
\item\label{kibble}
T.W.B. Kibble {\it J. Phys.} {\bf A9}, 1387 (1976).
\item\label{vachaspati}
T. Vachaspati, A. Vilenkin, \prl {\bf 67}, 1057 (1991).
\item\label{seitz}
S. Seitz, P. Schneider, J. Ehlers, \cqg {\bf 11}, 23 (1994).
\item\label{davis}
M. Davis, J. Huchra, D.W. Latham, J. Tonry \apj {\bf 253}, 423 (1982)
\item\label{barn} 
J. M. Barnothy, {\sl Astron. J.}, {\bf 70}, 666 (1965). 
\end{enumerate}
\vfill

\end{document}